# Comparison of inelastic and quasi-elastic scattering effects on nonlinear electron transport in quantum wires

Danhong Huang,[1] and Godfrey Gumbs[2]

[1] *Air Force Research Laboratory, Space Vehicles Directorate, Kirtland Air Force Base, NM 87117, USA and*

[2] *Department of Physics and Astronomy, Hunter College of the City University of New York, 695 Park Avenue, New York, NY 10065, USA*

## Abstract

When impurity and phonon scattering coexist, the Boltzmann equation has been solved accurately for nonlinear electron transport in a quantum wire. Based on the calculated non-equilibrium distribution of electrons in momentum space, the scattering effects on both the non-differential (for a fixed dc field) and differential (for a fixed temperature) mobilities of electrons as functions of temperature and dc field were demonstrated. The non-differential mobility of electrons is switched from a linearly increasing function of temperature to a parabolic-like temperature dependence as the quantum wire is tuned from an impurity-dominated system to a phonon-dominated one [see T. Fang, *et al.*, Phys. Rev. B **78**, 205403 (2008)]. In addition, a maximum has been obtained in the dc-field dependence of the differential mobility of electrons. The low-field differential mobility is dominated by the impurity scattering, whereas the high-field differential mobility is limited by the phonon scattering [see M. Hauser, *et al.*, Semicond. Sci. Technol. **9**, 951 (1994)]. Once a quantum wire is dominated by quasi-elastic scattering, the peak of the momentum-space distribution function becomes sharpened and both tails of the equilibrium electron distribution centered at the Fermi edges are raised by the dc field after a redistribution of the electrons is fulfilled in a symmetric way in the low-field regime. If a quantum wire is dominated by inelastic scattering, on the other hand, the peak of the momentum-space distribution function is unchanged while both shoulders centered at the Fermi edges shift leftward correspondingly with increasing dc field through an asymmetric redistribution of the electrons even in low-field regime [see C. Wirner, *et al.*, Phys. Rev. Lett. **70**, 2609 (1993)].

## I. INTRODUCTION

A clear understanding of the electron transport processes in both semiconductor quantum wells and quantum wires is of importance for the development of high-mobility quantum electronic devices. Several studies have been connected with the low-temperature electron conduction in quantum wires within the linear-transport regime.[1–12] The resonant effects of both magnetic field and electron-electron interaction on thermoelectric power in quantum wires were investigated.[10,13,14] In the presence of electron-electron interaction varying along quantum wires with a broken translational symmetry, the linear temperature dependence of resistivity was predicted.[15] When a voltage is applied between split gates, the backscattering of electrons at one of two rough edges inside a conduction channel was found enhanced experimentally, although the phenomenon has not been fully understood so far in a perpendicular magnetic field due to combined effects of the Lorentz force and a static potential across split gates.[16] For the effect of electron-phonon interaction on conduction carriers in quantum wires, non-Markovian Langevin-like equations were derived for calculating carrier temperature and a current-voltage instability was predicted.[17,18] Moreover, the Boltzmann transport equation was solved for linear transport of electrons in single and tunnel-coupled double quantum wires with a full inclusion of electron-phonon scattering on a microscopic level.[11]

In recent times, the linear magneto-quantum transport properties of electrons in single and tunnel-coupled double quantum wires have been extensively investigated by Lyo and Huang[8,11,19] based on an accurate solution of the linearized Boltzmann equation[20], in which the effects of both elastic and inelastic scattering as well as the effect of electron-electron scattering on the conductance of electrons had been treated microscopically beyond the well-known relaxation-time approximation[21,22]. Furthermore, the nonlinear electron transport under a strong dc field in a single quantum wire was also explored by using the coupled force balance[23] and microscopic scattering equations[24], in which only the elastic scattering effect on the electron mobility had been computed. The replacement of the energy-balance equation[25] for the relative scattering motion of electrons by the microscopic scattering equation allows one to incorporate the asymmetric electron distribution in momentum space.[24] However, the microscopic scattering equation still employs an adiabatic-type approximation[24–26] to the coupling between the center-of-mass and relative motions of electrons. The aim of present paper is to develop an efficient numerical method for solving accurately the Boltzmann transport equation which includes inelastic scattering beyond the linear-transport regime and further applying it to quantum -wire systems as a first step.

From the current study, we find that the coupled force balance and microscopic scattering equations [24] are still not accurate enough to describe the electron distribution as a function of momentum in the high-field regime. Although the calculated currents as functions of either dc-field or temperature can be compared directly with the experimentally measured ones, the momentum dependence of the electron distribution in the nonlinear-transport regime cannot be extracted reliably from the measured current on the microscopic level. Therefore, the accurate calculation of the steady-state non-equilibrium electron distribution based on the Boltzmann equation [20] becomes necessary for this purpose and is expected to provide a fair evaluation of the accuracy of the coupled force-balance and microscopic scattering equations in the high-field limit. In this paper, we modify the previously derived formalism by Huang, Lyo and Gumbs [27] for the Bloch oscillations of electrons within a quantum-dot superlattice to investigate the nonlinear transport of electrons in a single-subband quantum wire. Here, for completeness, we will include some of the key equations from our previous studies [27]. The destruction of phase coherence by electron-phonon scattering in disordered conductors has been investigated. [28] However, we shall confine our attention to high-quality quantum wires where the localization length is expected to be longer than the sample length. Compared with a strong electron-phonon interaction, the electron-electron interaction here is negligible [19] for a quantum-wire system with low electron densities. As long as the subband spacing is much larger than the average thermal energy as well as the Fermi energy of conduction electrons, the single-subband model employed here is justifiable. [8] This condition is easily achieved for a quantum-wire system with a strong lateral confinement and low electron densities at moderate to low lattice temperatures. As is well known, electron-electron interaction yields the Tomonaga-Lüttinger liquid effect in clean quantum-wire systems. Indeed, a conductance increase by electron-phonon interaction in quantum wires was predicted [29] for this case. However, this effect will not be considered here for diffusive electron transport.

Our main results in this paper include the comparisons of quasi-elastic with inelastic scattering effects on both the temperature $T$ and the dc-field $\mathcal{F}_0$ dependence of electron mobility, as well as on the wave number $k$-dependence of the electron extreme non-equilibrium distribution function. For the electron mobility as a function of $T$, its linear increase with $T$ in an impurity-dominated system [20] switches to a parabolic-like $T$ dependence in a phonon-dominated system with a maximum occurring at a characteristic temperature at which the strengths of quasi-elastic and inelastic scattering of electrons become comparable with each other. In the presence of scattering, we have found that the electron mobility increases (decreases) with $\mathcal{F}_0$ in the low (high) field regime, resulting in a maximum in the mobility as a function of $\mathcal{F}_0$. In addition, when the quantum wire radius $R$ is decreased, the characteristic temperature is also reduced due to the enhanced phonon scattering strength, and the maximum in the $\mathcal{F}_0$-dependent mobility shifts down to a lower $\mathcal{F}_0$ value. Furthermore, a well-known $1/T$ reduction of the mobility in the high-$T$ limit [20] is obtained when the electron density is lowered due to more accessible phonons contributing to a scattering process, and the mobility is increased (decreased) for low (high) $\mathcal{F}_0$ values because of the reduced impurity scattering (the enhanced nonlinear phonon scattering) of electrons, respectively.

For electron distribution as a function of $k/k_{\rm F}$ ($k_{\rm F}$ is the Fermi wave number), the peak at $k = 0$ is unchanged while both shoulders centered at $k = \pm k_{\rm F}$ shift leftward with increasing $\mathcal{F}_0$ in a quantum-wire system dominated by inelastic scattering. For a system dominated by quasi-elastic scattering, on the other hand, the peak of the distribution function at $k = 0$ becomes sharpened. At the same time, both tails of the equilibrium electron distribution in the region $|k| \geq k_{\rm F}$ are lifted up by the quasi-elastic scattering of electrons. Our calculated asymmetric (symmetric) distribution functions of electrons for low-field regime in the quantum-wire systems dominated by inelastic (quasi-elastic) scattering cannot be described by a quasi-equilibrium Fermi function employed in the energy balance equation with an effective electron temperature [25,26].

The paper is organized as follows. In Sec. II, we introduce our model for single-subband quantum wire systems, derive the general formalism for nonlinear electron transport, and solve the Boltzmann equation with the full inclusion of scattering by both impurities and acoustic phonons. In Sec. III, numerical results are presented for comparison between the effects for quasi-elastic and inelastic scattering of electrons on the temperature and dc-field dependence of mobility, as well as for a comparison of the quasi-elastic and inelastic scattering effects on the electron distributions in momentum space. The conclusions drawn from these calculated results are briefly summarized in Sec. IV.

## II. MODEL AND THEORY

In this section, we describe a microscopic approach for the Boltzmann equation in a quasi-one-dimensional quantum wire. Scattering by both impurities and acoustic phonons is fully taken into account. We assume a strong parabolic potential for the transverse confinement of the electrons.

We begin by considering a degenerate electron gas moving in a quantum wire in the $y$ direction, as shown in Fig. 1. For this system with a strong transverse confining potential, only the lowest subband will be occupied by electrons in the low-density and low-temperature regime, and therefore, a single subband model will be adequate. The original form of the Boltzmann equation for a time-dependent distribution function $f(k,\,t)$ of electrons in a quantum wire is given by [27]

$$\frac{\partial f(k,\,t)}{\partial t}=\frac{e\mathcal{F}_0}{\hbar}\,\frac{\partial f(k,\,t)}{\partial k}+\sum_{k'}\mathcal{P}_{\rm e}(k,\,k')\left[f(k',\,t)-f(k,\,t)\right]$$
$$+\sum_{k',\,\mathbf{q},\lambda}\left(\mathcal{P}_+(k,\,k')\,\left\{(n_{q\lambda}+1)\,f(k',\,t)\,[1-f(k,\,t)]-n_{q\lambda}\,f(k,\,t)\,[1-f(k',\,t)]\right\}\right.$$
$$\left.+\;\mathcal{P}_-(k,\,k')\,\left\{n_{q\lambda}\,f(k',\,t)\,[1-f(k,\,t)]-(n_{q\lambda}+1)\,f(k,\,t)\,[1-f(k',\,t)]\right\}\right)\ ,\ , \qquad (1)$$

where $\mathcal{F}_0$ denotes the applied dc electric field, $k$ is the electron wave vector along the $y$ direction, $\mathbf{q}$ is the three-dimensional wave vector of phonons, $\lambda=\ell$ ($\lambda=t$) labels the longitudinal (transverse) acoustic phonon modes, and $n_{q\lambda}=N_0(\omega_{q\lambda})$ is the Bose distribution function for the equilibrium phonons with frequency $\omega_{q\lambda}$ in the mode of $\mathbf{q}$ and $\lambda$. In Eq. (1), we have defined the electron-phonon inelastic scattering rate as [10,11,27]

$$\mathcal{P}_\pm(k,\,k')=\frac{2\pi}{\hbar}\,|V_{k,\,k'}|^2\,\delta(\varepsilon_k-\varepsilon_{k'}\pm\hbar\omega_{q\lambda})\,\delta_{k',\,k\pm q_y}\ , \qquad (2)$$

and the electron-impurity elastic scattering rate as [8,27]

$$\mathcal{P}_{\rm e}(k,\,k')=\frac{2\pi}{\hbar}\,|U_{k,\,k'}|^2\,\delta(\varepsilon_k-\varepsilon_{k'})\ , \qquad (3)$$

where $\varepsilon_k=\hbar^2k^2/2m^*$ is the electron kinetic energy with $m^*$ the electron effective mass, $V_{k,\,k'}$ is the interaction between electrons and phonons, $U_{k,\,k'}$ is the interaction between electrons and impurities, and the so-called $U$-scattering process [20] has been neglected due to low densities and low temperatures of the system.

We define $f(k,\, t) = f_k^{(0)} + g_k(t)$ with a dynamical non-equilibrium part $g_k(t)$ for the total distribution function, where $f_k^{(0)} = f_0(\varepsilon_k)$ is the Fermi function for equilibrium electrons. We note that the terms of zeroth order in $g_k$ in the scattering parts of Eq. (1) cancel out due to detailed balance. After incorporating these considerations, we can algebraically simplify Eq. (1) into

$$\begin{aligned}\frac{\partial g_k(t)}{\partial t} &= -e\mathcal{F}_0 v_k \left[-\frac{\partial f_k^{(0)}}{\partial \varepsilon_k}\right] + \frac{e\mathcal{F}_0}{\hbar}\,\frac{\partial g_k(t)}{\partial k} + \left[g_{-k}(t) - g_k(t)\right] \sum_{k' \in -k} \mathcal{P}_{\rm e}(k,\, k') \\ &- \; g_k(t) \sum_{k',\, \mathbf{q}, \lambda} \left\{\mathcal{P}_+(k,\, k') \left(n_{q\lambda} + f_{k'}^{(0)}\right) + \mathcal{P}_-(k,\, k') \left(n_{q\lambda} + 1 - f_{k'}^{(0)}\right)\right. \\ &+ \; g_{k'}(t) \left[\mathcal{P}_+(k,\, k') - \mathcal{P}_-(k,\, k')\right] \\ &+ \sum_{k',\, \mathbf{q}, \lambda} g_{k'}(t) \left[\mathcal{P}_+(k,\, k') \left(n_{q\lambda} + 1 - f_k^{(0)}\right) + \mathcal{P}_-(k,\, k') \left(n_{q\lambda} + f_k^{(0)}\right)\right] \; , \end{aligned} \tag{4}$$

where $v_k = (1/\hbar) d\varepsilon_k/dk = \hbar k/m^*$ is the group velocity of electrons with the wave number $k$ and the $k'$-sum for $k' \in -k$ is carried out near the resonance at $k' = -k$.

The transport relaxation rate from electron-impurity scattering equals[8]

$$\frac{1}{\tau_k} \equiv 2 \sum_{k' \in -k} \mathcal{P}_{\rm e}(k,\, k') = \frac{\pi}{\hbar} |U_{k,\, -k}|^2\, \mathcal{D}(\varepsilon_k) \; . \tag{5}$$

Here, the factor 2 arises from the $2k_F$ ($k_F$ is the Fermi wave vector at zero temperature) scattering across the Fermi sea and $\mathcal{D}(\varepsilon)$ is the total density-of-states given by

$$\mathcal{D}(\varepsilon) = \frac{L_0}{\pi\hbar} \sqrt{\frac{2m^*}{\varepsilon_k}} \; , \tag{6}$$

where $L_0$ is the length of the quantum wire. For electron-phonon scattering, we define scattering rates:

$$\mathcal{W}_\pm(k,\, k') \equiv \sum_{\mathbf{q},\, \lambda} \mathcal{P}_\pm(k,\, k') \; . \tag{7}$$

Making use of Eq. (2) and writing $\hbar\omega_{q\lambda} = \alpha_\lambda \sqrt{(k-k')^2 + q_\perp^2}$ from the Debye model, we have[11]

$$\mathcal{W}_\pm(k,\, k') = \frac{\mathcal{S}}{\hbar} \sum_\lambda \theta\left[(\varepsilon_k - \varepsilon_{k'})^2 - \alpha_\lambda^2 (k-k')^2\right]\, \theta(\pm\varepsilon_{k'} \mp \varepsilon_k)\, \frac{\hbar\omega_{q\lambda}\, |V_\lambda(\mathbf{q})|^2}{\alpha_\lambda^2}\, \Phi(q_\perp) \; , \tag{8}$$

where $\theta(x)$ is the unit step function, $\mathcal{S}$ is the sample cross-sectional area, $\alpha_\lambda = \hbar c_\lambda$, $c_\lambda$ is the sound velocity for the phonon in the $\lambda$ mode, $|q_y| = |k' - k|$,

$$q_\perp = \sqrt{q_x^2 + q_z^2} = \frac{1}{\alpha_\lambda}\sqrt{(\varepsilon_k - \varepsilon_{k'})^2 - \alpha_\lambda^2(k - k')^2}\ , \tag{9}$$

$$q = \sqrt{q_y^2 + q_\perp^2} = \frac{|\varepsilon_k - \varepsilon_{k'}|}{\alpha_\lambda}\ , \tag{10}$$

and the form factor in Eq. (8) takes the following form

$$\Phi(q_\perp) = \exp(-q_\perp^2 R^2/2) = \exp\left\{-[(\varepsilon_k - \varepsilon_{k'})^2 - \alpha_\lambda^2(k - k')^2]R^2/2\alpha_\lambda^2\right\} \tag{11}$$

for a transverse parabolic confining potential within the $xz$-plane with the radius $R$ for the cylindrical confinement model. The quantity $\Phi(q_\perp)$ arises from the consequence of momentum conservation for $q_\perp$. A smaller value of $R$ in Eq. (11) implies a stronger electron-phonon scattering in Eq. (8).

Assuming axial symmetry, the strength of the electron-phonon interaction $|V_\lambda(\mathbf{q})|^2$ in Eq. (8) is given by [10]

$$\hbar\omega_{q\lambda}\,|V_\lambda(\mathbf{q})|^2 = \frac{(\hbar\omega_{q\lambda})^2}{2[\epsilon(q_y)]^2\rho c_\lambda^2\mathcal{V}}\left[D^2\,\delta_{\lambda,\ell} + \left(\frac{eh_{14}}{q}\right)^2 A_{\mathbf{q}\lambda}\right]\ , \tag{12}$$

where $\mathcal{V} = \mathcal{S}L_0$ is the sample volume, $\rho$ is the mass density, $D$ is the deformation-potential coefficient, $h_{14}$ is the piezoelectric constant, $\epsilon(q_y)$ is the Thomas-Fermi dielectric function, and $A_{\mathbf{q}\lambda}$ is the structure factor for acoustic phonons due to anisotropic electron-phonon coupling. The quantum-wire structure factors of phonons in Eqs. (12) are calculated to be

$$A_{\mathbf{q}\ell} = \frac{9q_\perp^4 q_y^2}{2(q_\perp^2 + q_y^2)^3}\ , \qquad A_{\mathbf{q}t} = \frac{q_\perp^2(8q_y^4 + q_\perp^4)}{4(q_\perp^2 + q_y^2)^3}\ . \tag{13}$$

We further note from Eq. (8) that only one of $\mathcal{W}_\pm(k,\,k')$ is nonzero, i.e., either $\mathcal{W}_+(k,\,k') > 0$ for $\varepsilon_{k'} > \varepsilon_k$ or $\mathcal{W}_-(k,\,k') > 0$ for $\varepsilon_{k'} < \varepsilon_k$. This directly yields the result

$$\mathcal{W}_\pm(k,\,k') = \theta(\pm\varepsilon_{k'} \mp \varepsilon_k)\sum_\lambda \mathcal{W}_\lambda(k,\,k')\ ,\ , \tag{14}$$

where $\mathcal{W}_\lambda(k,\,k')=\mathcal{W}_\lambda(k',\,k)$ is given by

$$\mathcal{W}_\lambda(k,\,k')=\frac{\mathcal{S}}{\hbar}\,\theta\left[(\varepsilon_k-\varepsilon_{k'})^2-\alpha_\lambda^2(k-k')^2\right]\,|V_\lambda(\mathbf{q})|^2\,\frac{q}{\alpha_\lambda}\,\Phi(q_\perp)\ . \tag{15}$$

Using the above scattering rates, we rewrite Eq. (4) in the concise form

$$\begin{aligned}\frac{\partial g_k(t)}{\partial t}&=e\mathcal{F}_0v_k\,\frac{\partial f_k^{(0)}}{\partial\varepsilon_k}+\frac{e\mathcal{F}_0}{\hbar}\,\frac{\partial g_k(t)}{\partial k}+\frac{1}{2\tau_k}\,[g_{-k}(t)-g_k(t)]\\ &-\ g_k(t)\sum_{k'}\left\{\mathcal{W}_+(k,\,k')\left(n_{k,\,k'}+f_{k'}^{(0)}\right)+\mathcal{W}_-(k,\,k')\left(n_{k,\,k'}+1-f_{k'}^{(0)}\right)\right.\\ &+\ \left.g_{k'}(t)\,\left[\mathcal{W}_+(k,\,k')-\mathcal{W}_-(k,\,k')\right]\right\}\\ &+\ \sum_{k'}g_{k'}(t)\left[\mathcal{W}_+(k,\,k')\left(n_{k,\,k'}+1-f_k^{(0)}\right)+\mathcal{W}_-(k,\,k')\left(n_{k,\,k'}+f_k^{(0)}\right)\right]\ ,\end{aligned} \tag{16}$$

where $n_{k,\,k'}=N_0(|\varepsilon_k-\varepsilon_{k'}|/\hbar)$ is independent of $\lambda=\ell,\,t$. By introducing the notations $f_k^-=f_k^{(0)}$ and $f_k^+=1-f_k^{(0)}$, we obtain the discrete form for the total inelastic scattering rate

$$\mathcal{W}_j=\frac{L_0}{2\pi}\,\delta k\,\sum_{j',\,\pm}\mathcal{W}_\pm(k,\,k')\left(n_{k,\,k'}+f_{k'}^{\mp}\right), \tag{17}$$

with $k\equiv j\delta k$ and $k'\equiv j'\delta k$ $(\delta k\to 0)$, and

$$\mathcal{W}_{j,\,j'}=\sum_{\pm}\mathcal{W}_\pm(k,\,k')\left(n_{k,\,k'}+f_{k'}^{\pm}\right)\ . \tag{18}$$

We can also write

$$\mathcal{W}_j^g(t)=\frac{L_0}{2\pi}\,\delta k\,\sum_{j'}g_{k'}(t)\,\left[\mathcal{W}_+(k,\,k')-\mathcal{W}_-(k,\,k')\right]\ , \tag{19}$$

and

$$\begin{aligned}\frac{dg_k(t)}{dt}\ &=\ e\mathcal{F}_0v_k\,\frac{\partial f_k^{(0)}}{\partial\varepsilon_k}+\frac{e\mathcal{F}_0}{\hbar}\,\frac{\partial g_k(t)}{\partial k}\\ &-\ g_k(t)\,\left[\mathcal{W}_j+\mathcal{W}_j^g(t)\right]+\frac{1}{2\tau_k}\,[g_{-k}(t)-g_k(t)]+\frac{L_0}{2\pi}\,\delta k\,\sum_{j'}g_{k'}(t)\,\mathcal{W}_{j,\,j'}\ .\end{aligned} \tag{20}$$

In Eq. (20), the quantity $g_k\mathcal{W}_j^g$ is the only nonlinear term in $g_k$.

Equation (20) is equivalent to the following matrix equation for $1\le j\le N$

$$\frac{dg_j(t)}{dt}=b_j-\sum_{j'=1}^{N}a_{j,j'}(t)\,g_{j'}(t)\ , \tag{21}$$

where $k\equiv[j-(N+1)/2]\,\delta k$, $k'\equiv[j'-(N+1)/2]\,\delta k$, $\delta k=2k_{\rm max}/(N-1)$ with $N\gg 1$ an odd integer, and $|k|\leq k_{\rm max}$. The elements for the vector $\mathbf{b}$ in Eq. (21) are

$$b_j=e\mathcal{F}_0v_j\,\frac{\partial f_j^{(0)}}{\partial\varepsilon_j}\ , \tag{22}$$

and the elements for the matrix $\overleftrightarrow{\mathbf{a}}(t)$ in Eq. (21) are

$$\begin{aligned}a_{j,j'}(t)\ &=\ \delta_{j,j'}\left[\mathcal{W}_j+\mathcal{W}_j^g(t)+\frac{1-\delta_{j,(N+1)/2}}{2\tau_j}\right]-\delta_{j+j',N+1}\left[\frac{1-\delta_{j,(N+1)/2}}{2\tau_j}\right]\\ &-\ \frac{e\mathcal{F}_0}{\hbar\delta k}\,c_{j,j'}-\frac{L_0}{2\pi}\,\delta k\,\mathcal{W}_{j,j'}\ .\end{aligned} \tag{23}$$

By assuming that the dc electric field $\mathcal{F}_0$ is turned on at time $t=0$, the differential equation (21) can be solved in combination with the initial condition $g_j(0)=0$ for $j=1,\,2,\,\cdots,\,N$. In addition, using the so-called 3-point central-difference formula, which transforms the differential term $(e\mathcal{F}_0/\hbar)\,\partial g_k(t)/\partial k$ in Eq. (20) into a finite-difference term, we can write $c_{j,j'}$ in Eq. (23) in a simple form of

$$c_{j,j'}=\frac{1}{2}\left(\delta_{j,j'-1}-\delta_{j,j'+1}\right)\ . \tag{24}$$

Unfortunately, the quantities $\{g_1,\,g_2,\,\cdots,\,g_N\}$ are not linearly independent of each other, which implies that the matrix $\overleftrightarrow{\mathbf{a}}(t)$ in Eq. (21) is a singular one. In fact, the condition for the particle-number conservation requires that

$$\sum_{j=1}^{N}g_j(t)=0\ . \tag{25}$$

Combined with the fact that $\varepsilon_k=\varepsilon_{-k}$ and $v_k=-v_{-k}$, this yields for $1\leq j\leq N$

$$g_N(t)=-\sum_{j=1}^{N-1}g_j(t)\ , \tag{26}$$

where $j=N$ is the right-most $k$-space point. As a result of Eq. (26), we can renormalize the singular $(N\times N)$-matrix $\overleftrightarrow{\mathbf{a}}(t)$ in Eq. (21) into a regular $[(N-1)\times(N-1)]$-matrix $\overleftrightarrow{\mathbf{a}}'(t)$ through the following relation:

$$a'_{j,\,j'}(t) = a_{j,\,j'}(t) - a_{j,\,N}(t)\ , \tag{27}$$

where $j, j' = 1,\, 2,\, \cdots,\, N-1$. Consequently, Eq. (21) is renormalized into

$$\frac{dg'_j(t)}{dt} = b'_j - \sum_{j'=1}^{N-1} a'_{j,\,j'}(t)\, g'_{j'}(t)\ , \tag{28}$$

where the vectors $\mathbf{g}'(t)$ and $\mathbf{b}'$ are the same as the vectors $\mathbf{g}(t)$ and $\mathbf{b}$, respectively, without the last element with $j = N$. In a similar way, we correspondingly rewrite Eq. (19), by excluding the element with $j = N$, as

$$\mathcal{W}^g_j(t) = \frac{L_0}{2\pi}\,\delta k \sum_{j'=1}^{N-1} g'_{j'}(t)\ \{\mathcal{W}_+(j,\,j') - \mathcal{W}_-(j,\,j') - [\mathcal{W}_+(j,\,N) - \mathcal{W}_-(j,\,N)]\}\ . \tag{29}$$

Once the non-equilibrium part, $g'_j(t)$, of the total electron distribution function has been solved from Eq. (28), the transient drift velocity, $v_{\rm c}(t)$, of the system can be calculated according to

$$v_{\rm c}(t) = \left[\sum_{j=1}^{N} f^{(0)}_j\right]^{-1} \left\{\sum_{j=1}^{N-1} (v_j - v_N)\ g'_j(t)\right\}\ . \tag{30}$$

The steady-state drift velocity $v_{\rm d}$ of electrons is given by $v_{\rm c}(t)$ in the limit as $t \to \infty$. Furthermore, the steady-state conduction current is given by $I_{\rm c} = en_{\rm 1D}v_{\rm d}$, where $n_{\rm 1D}$ is the linear density of electrons in a quantum wire. The standard non-differential mobility for a fixed dc field is given by the ratio $\mu_{\rm nd} = v_{\rm d}/\mathcal{F}_0$, while the so-called differential mobility of electrons for a fixed temperature is defined through $\mu_{\rm d} = \partial v_{\rm d}/\partial\mathcal{F}_0$.

## III. NUMERICAL RESULTS AND DISCUSSIONS

For convenience of notation, we define $\tau_j = \tau_k$ at the energy $\varepsilon_j = \varepsilon_k$ in Eq. (5):

$$\frac{1}{\tau_j} = \gamma_0\sqrt{\frac{E_{\rm F}}{\varepsilon_j}}\ , \tag{31}$$

where $\gamma_0$ is the value of $1/\tau_j$ at $\varepsilon_j = E_{\rm F}$, and $E_{\rm F}$ is the chemical potential $\mu_0$ of electrons at $T = 0\,$K. For a fixed electron temperature $T$ and a linear density $n_{\rm 1D}$, $\mu_0$ is determined from the following relation:

$$n_{1\mathrm{D}} = \frac{\delta k}{\pi} \sum_{j=1}^{N} \frac{1}{\exp[(\varepsilon_j - \mu_0)/k_\mathrm{B}T] + 1} \ . \tag{32}$$

The Fermi wave vector is equal to $k_\mathrm{F} = \pi n_{1\mathrm{D}}/2$ and the Fermi energy is $E_\mathrm{F} = \hbar^2 k_\mathrm{F}^2/2m^*$ at $T = 0\,\mathrm{K}$. We chose GaAs as the host material in carrying out our numerical calculations and made use of the following parameters: the sound velocities $c_\ell = 5.14 \times 10^5\,\mathrm{cm/sec}$, $c_t = 3.04 \times 10^5\,\mathrm{cm/sec}$, the mass density $\rho = 5.3\,\mathrm{g/cm}^3$, the piezoelectric constant $h_{14} = 1.2 \times 10^7\,\mathrm{V/cm}$, and the deformation-potential coefficient $D = -9.3\,\mathrm{eV}$. For the elastic scattering, we chose $\gamma_0 = 1 \times 10^{12}\,\mathrm{s}^{-1}$. The other parameters, such as $R$, $\mathcal{F}_0$, $T$ and $n_{1\mathrm{D}}$, will be directly indicated in the figure captions.

### A. Mobility

For linear electron transport, the non-differential mobility $\mu_\mathrm{nd}$ only depends on the sample geometrical size, impurity and phonon scattering strengths, electron density $n_{1\mathrm{D}}$, and temperature $T$ but not on the dc field $\mathcal{F}_0$. On the other hand, the differential mobility $\mu_\mathrm{d}$ depends on $\mathcal{F}_0$ for nonlinear electron transport. In this case, the electron mobility approaches one constant at the low-field limit, while saturating at another constant in the high-field regime. Therefore, the $\mathcal{F}_0$-dependent electron mobility implies a nonlinear transport process of electrons in a system.

Figure 2 compares the quasi-elastic (very large values for $R$, blue dashed curves) with the inelastic ($R = 30$ Å, solid red curves) scattering effects for $n_{1\mathrm{D}} = 2 \times 10^5\,\mathrm{cm}^{-1}$ on the $T$ dependence of non-differential mobility [in (a) when $\mathcal{F}_0 = 40\,\mathrm{V/cm}$] defined by $\mu_\mathrm{nd} = v_\mathrm{d}/\mathcal{F}_0$, as well as on the $\mathcal{F}_0$ dependence of differential mobility [in (b) with $T = 20\,\mathrm{K}$] defined by $\mu_\mathrm{d} = \partial v_\mathrm{d}/\partial \mathcal{F}_0$. For the $T$ dependence in Fig. 2(a), $\mu_\mathrm{nd}$ exhibits a simple linear relation when only the quasi-elastic scattering (i.e., with a very large value for $R$) exists. However, after the inelastic electron scattering has been introduced to the system, $\mu_\mathrm{nd}$ decreases strongly, and the straight line is replaced by a parabola-like curve with a maximum occurring at a characteristic temperature. The characteristic temperature corresponds to the case when the strengths of both quasi-elastic and inelastic scattering become comparable to each other. Similar thermal behavior was predicted by Fang, *et al.*, [30] for the linear electron transport in graphene nanoribbons. On the other hand, for $\mu_\mathrm{d}$ as a function of $\mathcal{F}_0$ in Fig. 2(b), we find $\partial\mu_\mathrm{d}/\partial\mathcal{F}_0 > 0$ for the small values of $\mathcal{F}_0$ but $\partial\mu_\mathrm{d}/\partial\mathcal{F}_0 < 0$ for high $\mathcal{F}_0$ when either the quasi-elastic or the inelastic scattering of electrons is present. A strong reduction in $\mu_\mathrm{d}$ at low $\mathcal{F}_0$ is found for phonon scattering, however, this difference becomes much smaller at high $\mathcal{F}_0$.

In Fig. 3, we display the inelastic scattering effects at $n_\mathrm{1D} = 2\times10^5\,\mathrm{cm}^{-1}$ on the $T$ dependence of $\mu_\mathrm{nd}$ [in (a) with $\mathcal{F}_0 = 40\,\mathrm{V/cm}$] and $\mathcal{F}_0$ dependence of $\mu_\mathrm{d}$ [in (b) with $T = 20\,\mathrm{K}$] for electrons in quantum wires with $R = 10$ Å (solid red curves) and $R = 30$ Å (blue dashed curves). When $R$ is decreased from 30 Å to 10 Å in Fig. 3(a), i.e. the phonon-scattering strength is increased, the system dominated by the quasi-elastic-scattering with $\partial\mu_\mathrm{nd}/\partial T > 0$ for small values of $T$ switches to one dominated by the inelastic-scattering with $\partial\mu_\mathrm{nd}/\partial T < 0$ because of the decrease of the characteristic temperature with decreasing $R$. For the $\mathcal{F}_0$ dependence of $\mu_\mathrm{d}$ in Fig. 3(b), the dc-field value corresponding to the maximum in $\mu_\mathrm{d}$ is shifted downwards as the radius $R$ of the quantum wire is decreased to 10 Å, in combination with a strong reduction in $\mu_\mathrm{d}$ at all the field values shown in this figure.

From Figs. 2(b) and 3(b) we find that $\mu_\mathrm{d}$, as a function of $\mathcal{F}_0$, initially increases but eventually decreases with $\mathcal{F}_0$ when the inelastic scattering effect is not very strong. In this case, the frictional force acted on electrons can be reduced [25,31,32] as long as the lattice temperature $T$, as well as the impurity density, are low and the electron drift velocity $v_\mathrm{d}$ is less than the sound velocity $c_\lambda$. The reduced frictional force to moving electrons is associated with the softened acoustic-phonon mode due to the Doppler shift. However, $\mu_\mathrm{d}$ of hot electrons decreases significantly once the nonlinear phonon scattering becomes dominant with increasing $\mathcal{F}_0$.

Figure 4 presents the results when inelastic scattering effects are present for $R = 10$ Å on the $T$ dependence of $\mu_\mathrm{nd}$ [in (a) with $\mathcal{F}_0 = 40\,\mathrm{V/cm}$] and the $\mathcal{F}_0$-dependence of $\mu_\mathrm{d}$ [in (b) with $T = 20\,\mathrm{K}$] for the quantum wires with $n_\mathrm{1D} = 2\times10^5\,\mathrm{cm}^{-1}$ (solid red curves) and $n_\mathrm{1D} = 1\times10^5\,\mathrm{cm}^{-1}$ (dashed blue curves). When $n_\mathrm{1D}$ decreases from $2\times10^5\,\mathrm{cm}^{-1}$ to $1\times10^5\,\mathrm{cm}^{-1}$ in Fig. 4(a), more acoustic phonons in the system become accessible for the inelastic scattering of electrons since $\omega_{q\lambda} \sim 2c_\lambda k_\mathrm{F} \propto n_\mathrm{1D}$ in this case. As a result of this, the well-known $1/T$ reduction of $\mu_\mathrm{nd}$ at the high-temperature limit [20] is seen. At the same time, $\mu_\mathrm{nd}$ is enhanced for low $T$ but reduced for high $T$ with decreasing $n_\mathrm{1D}$. The enhancement of $\mu_\mathrm{nd}$ at low $T$ is due to the reduced $1/\tau_j \propto \sqrt{E_\mathrm{F}} \propto n_\mathrm{1D}$ in Eq. (31) for the quasi-elastic scattering of electrons, while the reduction of $\mu_{\rm nd}$ at high $T$ directly comes from increased phonon scattering proportional to $T$ once $k_{\rm B}T \gg \hbar\omega_{q\lambda}$. A small enhancement in $\mu_{\rm d}$ for low $\mathcal{F}_0$ and a large reduction in $\mu_{\rm d}$ for high $\mathcal{F}_0$ are also found in Fig. 4(b) as $n_{\rm 1D}$ is reduced to $1\times 10^5\,{\rm cm}^{-1}$. In this case, however, the dc-field value corresponding to the maximum in $\mu_{\rm d}$ is not shifted. The little enhancement of $\mu_{\rm d}$ at low $\mathcal{F}_0$ is attributed to the decreased quasi-elastic scattering of electrons, while the significant reduction of $\mu_{\rm d}$ at high $\mathcal{F}_0$ is a consequence of the increased nonlinear phonon scattering [$\mathcal{W}^g_j(t)$ in Eq. (29)] which directly contributes to the $\mathcal{F}_0$ dependence of $\mu_{\rm d}$ for nonlinear electron transport. In this case, the low-field mobility is dominated by impurity scattering, while the high-field mobility is limited by phonon scattering. The increase in the high-field mobility with electron density was experimentally observed previously by Hauser, *et al.*, [33] in a quantum wire.

### B. Distribution Function

From Eq. (30) we are aware that the steady-state drift velocity $v_{\rm d}$ is uniquely determined by the non-equilibrium part $g'_k(t)$ of the total electron distribution function $f(k,\,t)$ in the limit $t\to\infty$. From a physical point of view, this implies that there must be a microscopic origin for all the observed behaviors for nonlinear electron transport in Figs. 2(b), 3(b) and 4(b), and this origin is believed to be related to the field-induced redistribution of conduction electrons in $k$-space.

In Fig. 5, we show the inelastic scattering effects for $R = 10$ Å, $T = 10\,{\rm K}$ and $n_{\rm 1D} = 2\times 10^5\,{\rm cm}^{-1}$ in both the non-equilibrium part of electron distribution function $g_k$ [in (a)] and the total electron distribution function $f_k$ [in (b)] as functions of a scaled electron wave number $k/k_{\rm F}$ for $\mathcal{F}_0 = 10\,{\rm V/cm}$ (black solid curves), $20\,{\rm V/cm}$ (red dashed curves), $40\,{\rm V/cm}$ (green dash-dotted curves), and $60\,{\rm V/cm}$ (blue dash-dot-dotted curves). For the sake of comparison, we also plot the Fermi function $f^{(0)}_k$ (thick black dashed curve) in Fig. 5(b). In Fig. 5(a), $g_k > 0$ ($g_k < 0$) represents the population (depopulation) of $k$-state electrons in the quantum wire. When $\mathcal{F}_0$ is gradually increased from $10\,{\rm V/cm}$ up to $60\,{\rm V/cm}$, the inelastic phonon scattering in the system moves more and more electrons from the right Fermi edge at $k/k_{\rm F} = +1$ to a high-energy states peaked at $k/k_{\rm F} = -2$ after passing through the left Fermi edge at $k/k_{\rm F} = -1$. However, the electrons at the lowest $k = 0$ energy state are left largely untouched. This significant adjustment of conduction electron distribution in $k$-space can be also seen from the total distribution function $f_k$ in Fig. 5(b), where the peak at $k = 0$ is unchanged while both shoulders centered at $k/k_F = \pm 1$ are shifted to the left with increasing $\mathcal{F}_0$. This leads to an asymmetric distribution function with respect to $k = 0$ even in the low-field regime, a non-quasi-equilibrium electron distribution. A similar Fermi-edge shifting in a hot-electron distribution with an increasing dc field was experimentally observed by Wirnel, *et al.*, [34] in GaAs/AlGaAs heterostructures.

Figure 6 exhibits the effects of quasi-elastic electron scattering at $T = 10\,$K and $n_{1D} = 2 \times 10^5\,\text{cm}^{-1}$ on $g_k$ [in (a)] and $f_k$ [in (b)] as functions of $k/k_F$ for $\mathcal{F}_0 = 10\,$V/cm (black solid curves), 20 V/cm (red dashed curves), 40 V/cm (green dash-dotted curves), and 60 V/cm (blue dash-dot-dotted curves). For clarity, we also include $f_k^{(0)}$ (thick black dashed curve) in Fig. 6(b). As $\mathcal{F}_0$ increases from 10 V/cm up to 60 V/cm in Fig. 6(a), all the conduction electrons in the low-energy $k$ states between the left and right Fermi edges are depopulated and moved to high-energy $k$ states outside the region with $|k/k_F| \leq 1$. However, the depopulation of electrons at $k = 0$ is found to be smaller than those at the two Fermi edges. This specific redistribution of electrons in $k$-space by quasi-elastic scattering is also reflected in Fig. 6(b) for $f_k$, where the peak at $k = 0$ is sharpened and the electrons around the peak are moved outward to the high-energy tail states of the equilibrium distribution function $f_k^{(0)}$ in $k$-space. However, moving electrons outward by quasi-elastic scattering in both directions in $k$-space creates a somewhat symmetric distribution of electrons in the low-field regime with a sharpened peak and two raised tails at the same time, in strong contrast to the result for inelastic scattering in Fig. 5(b). Although more and more high-energy states become populated by moving electrons from low-energy states through both quasi-elastic scattering and inelastic scattering of electrons, moved electrons in high-energy states only have negative group velocities, i.e. $d\varepsilon_k/dk < 0$, for the inelastic scattering case.

## IV. CONCLUDING REMARKS

The accurate solution of the Boltzmann equation for a quantum wire has been obtained with the full inclusion of both impurity and phonon scattering for conduction electrons on the microscopic level. Using this solution, both the non-differential and differential mobilities of the electrons as functions of temperature and dc field, as well as the non-equilibrium distribution of electrons in momentum space, were analyzed.

As a result of our studies, we have found that the non-differential mobility of electrons in a quantum wire switches from a linear increase with $T$ in an impurity-dominated system to a parabolic-like $T$ dependence in a phonon-dominated system. The characteristic temperature, at which the inelastic and quasi-elastic scattering strengths become comparable to each other, corresponds to a maximum in the non-differential mobility as a function of temperature, and it shifts with the quantum-wire radius. We also find that the differential mobility of electrons has a maximum as a function of dc field, and the field corresponding to this maximum shifts as the quantum-wire radius is varied. The low-field mobility is dominated by impurity scattering, while the high-field mobility is limited by phonon scattering.

For a quantum-wire system dominated by quasi-elastic scattering, the peak in the distribution function at $k = 0$ becomes sharpened, and both tails of the equilibrium electron distribution centered at the Fermi edges are lifted up by the dc field. On the other hand, for a quantum wire dominated by inelastic scattering, the peak at $k = 0$ is unchanged, while both shoulders centered at the Fermi edges shift to the left correspondingly with increasing dc field. Therefore, the quasi-elastic scattering in a quantum wire redistributes the conduction electrons in a symmetric way in momentum space in the low-field regime. However, the inelastic scattering of electrons redistributes them in an asymmetric way even in the low-field regime.

**Acknowledgments**

The authors would like to thank the Air Force Office of Scientific Research (AFOSR) for its support.

---

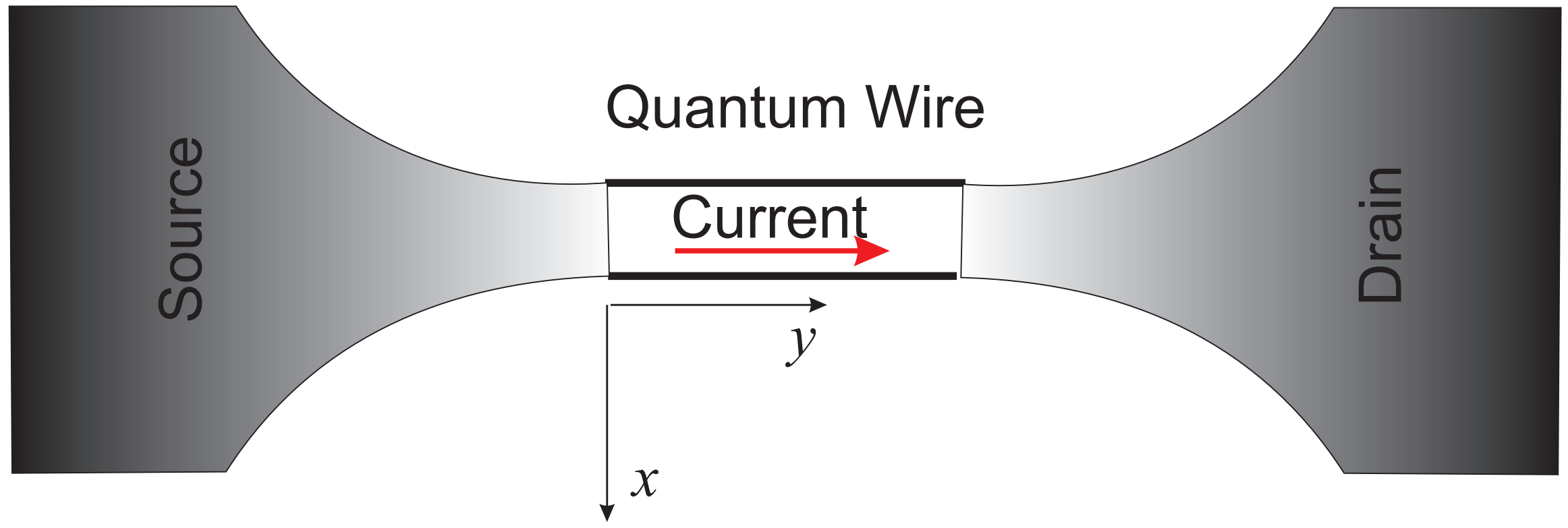

FIG. 1: (Color online) Schematic illustration for a quasi-one-dimensional quantum wire driven by an electric field along the $y$ (wire) direction, where the confinements in the $z$ and $x$ directions are given by a cylindrical parabolic potential. The current flowing through the channel between source and drain electrodes is indicated by a horizontal (red) arrow.

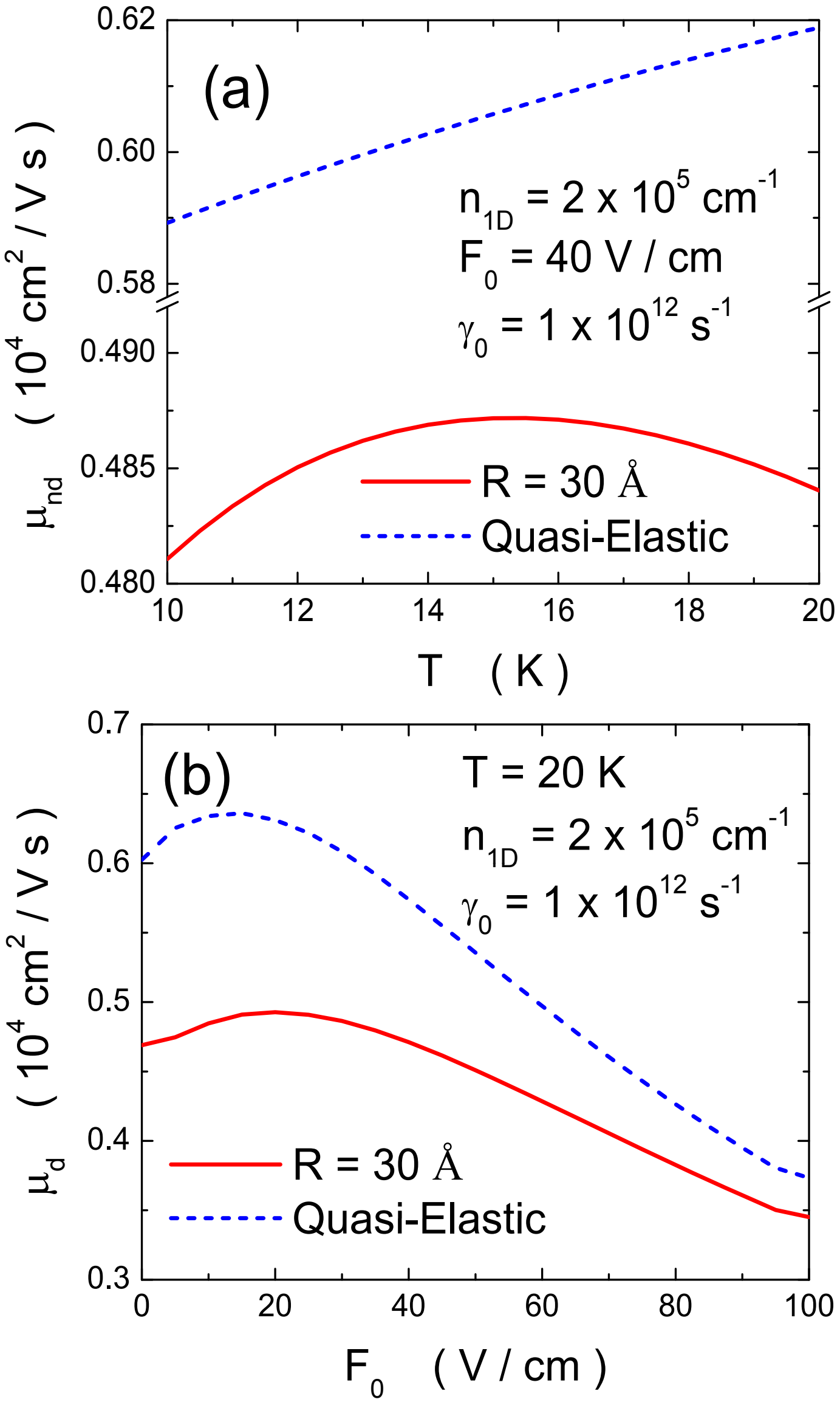

FIG. 2: (Color online) Calculated non-differential mobility $\mu_{\rm nd} = v_{\rm d}/\mathcal{F}_0$ in (a) as a function of temperature $T$ and differential mobility $\mu_{\rm d} = \partial v_{\rm d}/\partial\mathcal{F}_0$ in (b) as a function of dc-field $\mathcal{F}_0$ for inelastic scattering with $R = 30$ Å, (solid red curves) and quasi-elastic scattering (blue dashed curves) at $n_{1\rm D} = 2\times10^5\,{\rm cm}^{-1}$. In (a), we take $\mathcal{F}_0 = 40\,{\rm V/cm}$, while we choose $T = 20\,{\rm K}$ in (b).

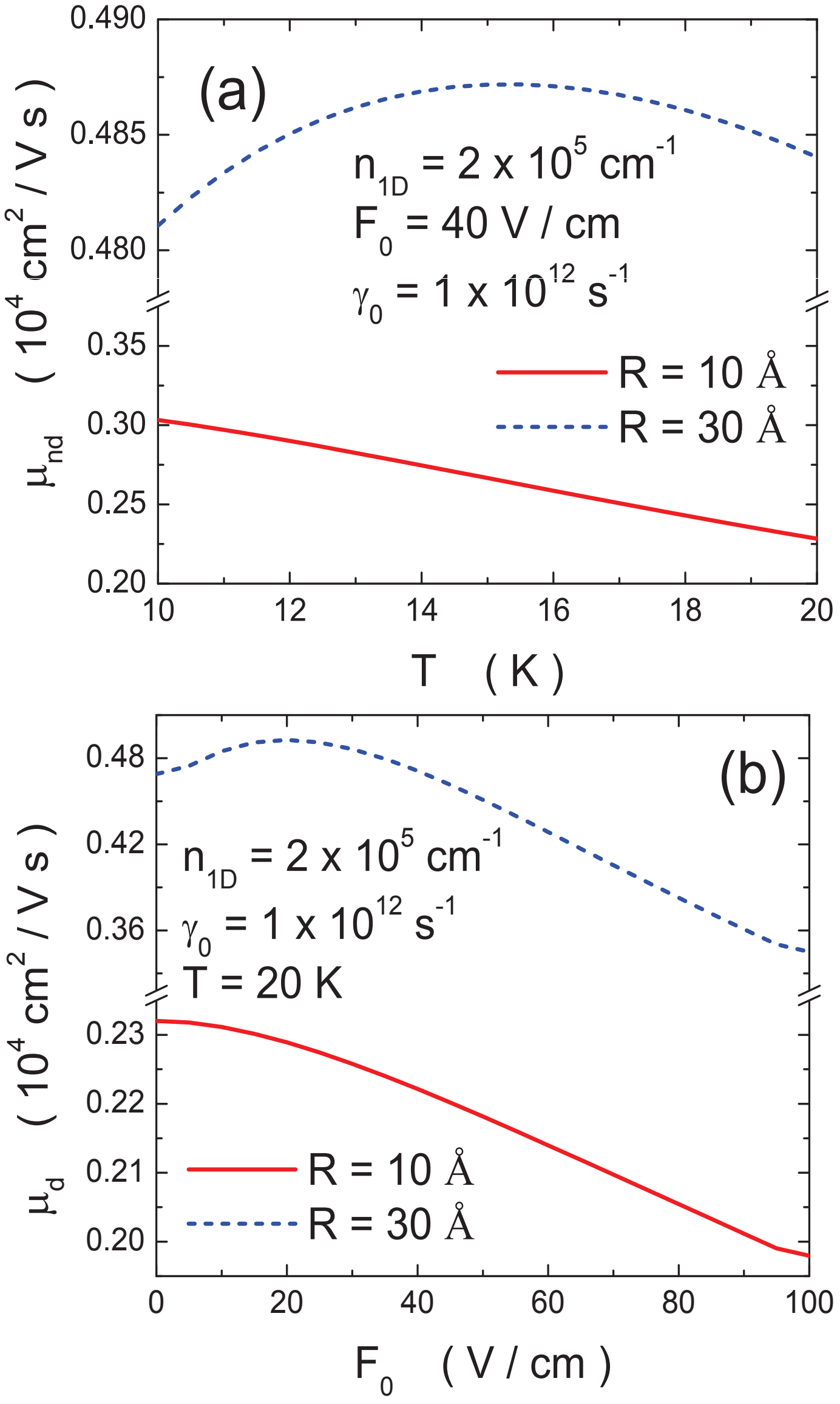

FIG. 3: (Color online) Numerical values of $\mu_{\rm nd}$ in (a) as a function of $T$ and $\mu_{\rm d}$ in (b) as a function of $\mathcal{F}_0$ for two values of $R = 10$ Å, (solid red curves) and $R = 30$ Å (blue dashed curves) at $n_{\rm 1D} = 2\times 10^5\,{\rm cm}^{-1}$. In (a), we take $\mathcal{F}_0 = 40\,{\rm V/cm}$, while we choose $T = 20\,{\rm K}$ in (b).

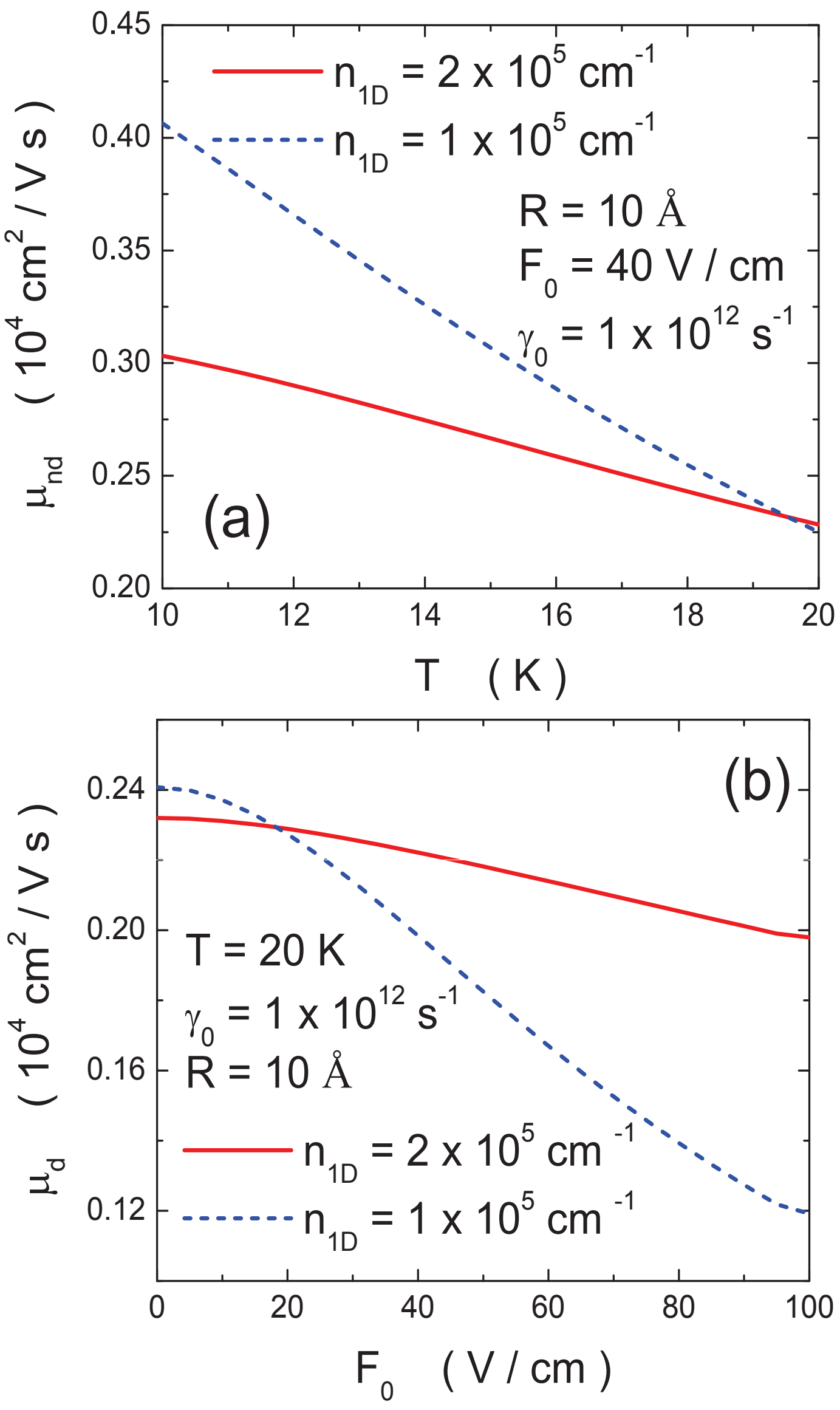

FIG. 4: (Color online) Calculated $\mu_{\rm nd}$ in (a) as a function of $T$ and $\mu_{\rm d}$ in (b) as a function of $\mathcal{F}_0$ for different electron densities $n_{\rm 1D} = 2 \times 10^5\,{\rm cm}^{-1}$ (solid red curves) and $n_{\rm 1D} = 1 \times 10^5\,{\rm cm}^{-1}$ (blue dashed curves) with $R = 10$ Å. In (a), we take $\mathcal{F}_0 = 40\,{\rm V/cm}$, while we choose $T = 20\,{\rm K}$ in (b).

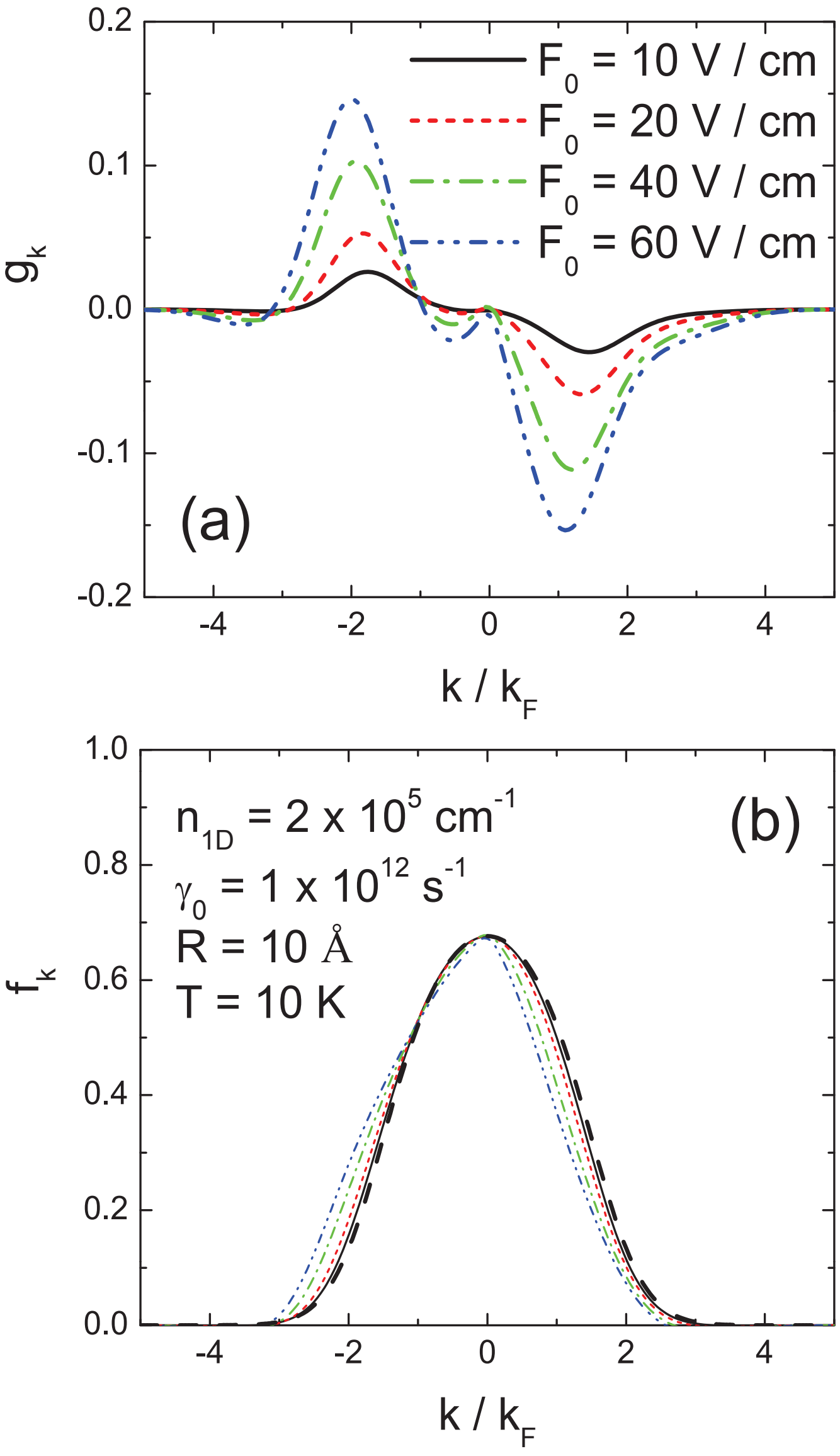

FIG. 5: (Color online) Steady-state $g_k$ in (a) and $f_k$ in (b) as functions of $k/k_{\rm F}$ for various dc-fields: $\mathcal{F}_0 = 10\,$V/cm (black solid curves), 20 V/cm (red dashed curves), 40 V/cm (green dash-dotted curves), and 60 V/cm (blue dash-dot-dotted curves). Here, we take $R = 10$ Å, $T = 10\,$K and $n_{\rm 1D} = 2 \times 10^5\,{\rm cm}^{-1}$. For the sake of comparison, we also show $f_k^{(0)}$ (thick black dashed curve) in (b).

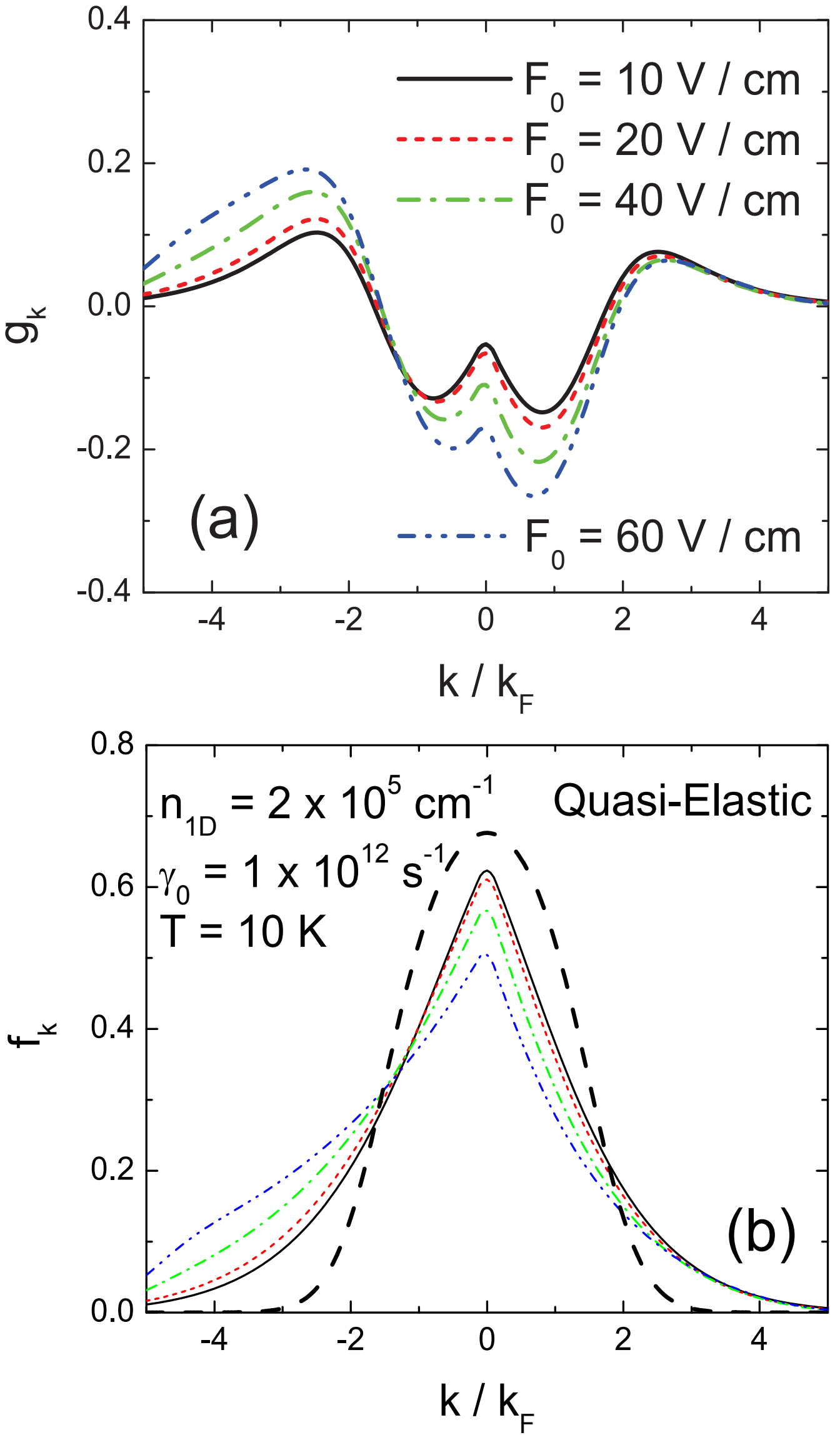

FIG. 6: (Color online) Plots of steady-state $g_k$ in (a) and $f_k$ in (b) as functions of $k/k_\mathrm{F}$ for various dc-fields: $\mathcal{F}_0 = 10\,\mathrm{V/cm}$ (black solid curves), $20\,\mathrm{V/cm}$ (red dashed curves), $40\,\mathrm{V/cm}$ (green dash-dotted curves), and $60\,\mathrm{V/cm}$ (blue dash-dot-dotted curves). Here, we take $T = 10\,\mathrm{K}$, $n_\mathrm{1D} = 2\times10^5\,\mathrm{cm}^{-1}$ and include the quasi-elastic scattering. For a clear comparison, we also display $f_k^{(0)}$ (thick black dashed curve) in (b).